\documentclass[pra,amsmath,amssymb]{revtex4}

\usepackage{epsfig}

\usepackage{color}
\usepackage{ulem}

\begin{document}

\title{ Finite Linear Spaces , Plane Geometries, Hilbert spaces and Finite Phase Space.}

\author{M. Revzen and A. Mann}
\affiliation {Department of Physics, Technion - Israel Institute of Technology,
Haifa 32000, Israel}

\date{\today}

\begin{abstract}

Finite plane geometry  is associated with finite dimensional Hilbert space.
 The
association allows mapping of q-number Hilbert space observables
 to the c-number formalism of quantum mechanics
in phase space. The mapped entities reflect geometrically based line-point
interrelation. Particularly simple formulas are involved when
use is made of mutually unbiased bases (MUB) representations for the Hilbert space entries.\\

The geometry specifies a point-line interrelation. Thus  underpinning
d-dimensional Hilbert space operators (resp. states) with geometrical points
leads to operators termed "line operators" underpinned by the
geometrical lines. These "line operators", $\hat{L}_j;$ (j designates the
line) form a complete orthogonal basis for Hilbert space operators. The
representation of Hilbert space operators in terms of these operators form the
phase space representation of the d-dimensional Hilbert space.\\

Examples for the use of the "line operators" in mapping (finite dimensional)
Hilbert space operators onto finite dimensional phase space functions are
considered. These include finite dimensional Wigner function and Radon
transform and a geometrical interpretation for the involvement of parity in
the mappings of Hilbert space onto phase space.\\

Two d-dimensional particles {\it product states} are underpinned with
geometrical points. The states, $|L_j\rangle$ underpinned with the
corresponding geometrical lines are maximally entangled states (MES). These
"line states" provide a complete $d^2$ dimensional orthogonal MES basis for
for the two d-dimensional particles.\\

The complete $d^2$ dimensional MES i.e. the "line states" are shown to provide
a transparent geometrical interpretation to the so called Mean King Problem
and its variant.\\

The "line operators" (resp. "line states") are studied in detail. \\

The paper aims at self sufficiency and to this end all relevant notions
are explained herewith.\\

\end{abstract}

\maketitle

\section{Introduction}

Phase space formulation of continuum (i.e. pertaining to Hilbert space
dimensionality $d\rightarrow \infty$) quantum mechanics was initiated with
\cite{weyl,wigner} and developed into a coherent autonomous approach to
quantization by \cite{groenewold, moyal, cohen}. The formalism  was (and is)
clarified and developed by numerous workers (cf. references in,
e.g.,\cite{schroeck,zachos}). It finds wide use in quantum optics
\cite{schleich,glauber,klauder, sudarshan,perelomov}, quantum cryptography
\cite{ulf} and foundation of quantum mechanics \cite{schroeck,zachos,bracken}.
It may be epitomized, perhaps, by Glauber's coherent state
\cite{glauber,klauder,sudarshan} which relates directly to phase space.\\

 The present study is concerns with finite dimensional phase space. It deals
with mapping finite dimensional
 Hilbert space onto (finite dimensional) phase space
 \cite{schwinger,wootters1, wootters2,klimov1, vourdas, tal, durt, rev1}. These
mappings are  epitomized, to
 a large extent, with wave functions of mutually unbiased bases
 (MUB) ( a brief review of MUB is given in Section III below). Physically these
states  encapsulate complementarity
- a fundamental quantum mechanical feature \cite{durt, bengtsson,combescure}.
Mathematically MUB were related to algebraic (Galois) fields \cite{tal, durt,
bengtsson,saniga}.  Of special interest in the present work is  a deep relation, first noted in
\cite{saniga} and \cite{wootters4}, between (finite) affine plane geometry (APG) and MUB. Such a
relation is implied by the known (e.g.\cite{bennett}) relation between
algebraic fields and geometry.  Thus finite dimensional algebraic fields (Galois fields,GF )
were (and are) used
extensively to extend analysis of unitary bases in finite
dimensional Hilbert space \cite{schwinger} to finite dimensional phase space quantum
mechanics \cite{ivanovich,tal, wootters1,wootters2,gibbons, klimov1,rev1}.
In the "triangle" GF- MUB - APG the emphasis of
the present study is on the MUB - APG side. More specifically we study the interrelation
between the dual of APG, viz. DAPG and MUB. (both APG and DAPG are reviewed in Section II.) For example,
 we show \cite{wootters4, klimov1, vourdas,tomer,rev1,rev2} in Section IV, that letting
the points of the geometry underpin MUB
state projectors -  the
physical entities (i.e. the Hilbert space operators) that the corresponding geometrical lines
underpin, dubbed "line operators" $\hat{L}_j$ (j designates a line)
form a complete orthonormal basis for \textcolor{blue}{ operators} in the d-dimensional
(Hilbert) space
understudy. The expansion coefficients (i.e. the representation) of Hilbert space
operators
in terms of these "line operators" are the finite dimensional phase space mappings of
the operators.
In particular the representation of the density matrix, $\rho$, in the space of these
 $\hat{L}_j$
is the finite dimensional Wigner  function \cite{wootters1, klimov1,vourdas,tomer} :
lines being parametrized by $j=\ddot{m};m_0$ which corresponds to phase space coordinates q,p in
the continuum \cite{rev1,khanna,pier,ady}.\\

The paper is organized as follows.
The succeeding section, Section II, contains  a explanatory discussion of the basic notions of
linear spaces and outline the approach \cite{batten}, that we adopt, of viewing geometries  as
constrained linear spaces, {\cal{S}}. The section, Section II, contains the definitions and
postulates of finite geometry \cite{bennett,batten} and the definition of
the corresponding interrelations among the Hilbert space entities underpinned by the geometry.\\

Section III presents the definitions and essential features of mutual unbiased bases
 (MUB) and collective coordinates in finite dimensional Hilbert spaces \cite{r1} that
are convenient for the labeling of Hilbert space {\it states} underpinned
by the geometry. Maximally entangled states are shown to form "line states" viz are underpinned by
the geometrical lines where the geometrical points underpin two particles product MUB states.\\
For a d-dimensional Hilbert space the maximal number of MUB is d+1 \cite{ivanovich,wootters1,durt}.
However  realization of d+1 MUB is known only for $d=p^m$, p a prime and m a (positive) integer. i.e.
precisely for the order for which (finite) algebraic (Galois) fields are known to exist, and precisely
the order for which
(finite) projective (and both affine and dual affine discussed in Section II) geometries exist. (There is
no mathematical proof for the exclusiveness of $d=p^m$ as the dimensionalities within which MUB and the
geometries exist. However no counter example is known \cite{bennett,durt}.) Thus it is  possible to
construct a geometry of order d given
that an algebraic field of this order exist. Indeed it was conjectured in \cite{saniga1} that "the
existence of d+1 MUB for d dimensional Hilbert space, {\it if d differs from a power of prime} is
intimately linked with the existence
of projective planes of this order. The present work is limited to the simplest cases, viz d=prime
$\ne 2$.
For these cases MUB and the geometries are realizable with relative ease. The extension to d=$p^m$,
i.e. power of prime, including 2, is possible and is briefly discussed in an appendix.\\

The next section, Section IV, accounts the actual underpinning of MUB projectors
and two particle states with finite geometry. It is argued that the most convenient
geometry for underpinning MUB entities is the dual affine plane geometry (DAPG)
which is discussed in detail. \\

In Section V mappings of (selected) Hilbert space entities onto finite dimensional
phase space is presented. We give here the derivation of finite dimensional Wigner
function, Radon transform and the parity operator as c number function in phase space.\\

d-dimensional Hilbert space may accommodate d+1 mutually unbiased bases (MUB) \cite{ivanovich,durt,tal}.
The existence of the d+1
bases is known only for $d=p^m$ (p a prime and m a positive integer). The order, d, referred to as
dimension in this paper, for which finite affine plane geometry (APG) and it dual (DAPG) are known to
exist is likewise $d=p^m$ \cite{bennett}.  The analysis of this paper is confined to d=p (a prime) and
extension to power of prime is only briefly discussed.\\

Some use is made of mathematically known results that are not studied in detail. The three
appendices are our struggle  at providing a descriptive account of the most important of these.
Thus Appendix A illustrates the idea of field extension. In Appendix B we considers a case where
few attributes of a linear space allows the deduction of all the essential characteristics of
a geometry. Appendix C provides a proof of the equality of the number of pencils in DAPG (the notions
involved
are elaborated on in Section II) with the number of points on a DAPG line.\\

\section{Finite Geometries, Linear spaces}

We begin with the definition of linear space $\cal{S}$: $\cal{S}$ is made of pair
entities, points and lines. There are $\nu$ points,
$S_{\alpha};\;\;\alpha=1,2...,\nu$ and there are $\cal{B}$ sets of points termed lines
$L_j;\;\;j=1,2...,\cal{B}$. These are interrelated via the following axioms:\\

$\lambda_1$. Given two distinct points there is exactly one line common to
both. Every point is common to at least two lines. \\

$\lambda_2$. There are at least two distinct points in a line. There are three
non co-linear points.\\

Correspondingly we define dual linear space $\tilde{\cal{S}}$ which is made up
of the same entities but their interrelation is gotten by exchanging  $points
\leftrightarrow lines$. Thus the axioms for $\tilde{\cal{S}}$ are:\\

$\tilde{\lambda}_1$. Given two distinct lines there is exactly one point
common to both. Every line is common to at least two points.\\

$\tilde{\lambda}_2$. There are at least two distinct lines with a common
point.
There are three lines with no three fold common point.\\

Our interest is with constrained linear spaces: Adding a third axiom, A, to $\lambda_1,\lambda_2$
 defines a constrained linear space $\cal{A}$:\\

A: Given a line L and a point $S_{\alpha}$ not on L, there exists exactly one
line, L', containing $S_{\alpha}$ such that $L \bigcap L'= \emptyset$. i.e.
$L'\parallel L.$\\

The three axioms $\lambda_1,\lambda_2\; and  \;A$ \textcolor{blue}{ define (finite)
affine plane geometry, APG} \cite{bennett}, i.e. ${\cal{A}} \equiv$ APG.\\

Correspondingly, adding a third axiom $\tilde{A}$ to those of the dual linear space above,
$\tilde{\cal{S}}$ defines  the linear space $\tilde{\cal{A}}$ :\\

$\tilde{A}$: Given a point $S_{\alpha}$ and a line L not containing the point
there exists exactly one point  $S_{\alpha{'}}$ on L with no line containing
them both.\\

The three axioms $\tilde{\lambda}_1,\;\tilde{\lambda}_2\; and \;\tilde{A}$ defines
 \textcolor{blue}{dual affine plane geometry
(DAPG)}, i.e. $\tilde{\cal{A}}\equiv$ DAPG.\\

In addition to the above two linear spaces, $\cal{A}$ - the APG and
$\tilde{\cal{A}}$- the DAPG , we consider a third linear space, $\cal{P}$ -
\textcolor{blue}{ the (finite) projective geometry, FPP}. This is
defined by constraining $\lambda_1,\lambda_2$ with \cite{batten}\\

P1. Any two distinct lines have a point in common.\\

P2. There are four points, no three of which are on common line.\\

\textcolor{blue}{Thus the linear space $\cal{P}$- FPP - is defined by $\lambda_1,\lambda_2, P1, P2.$}\\

Some general mathematically proved results for finite geometries will now be listed. $\nu$ is the number of
points and $\cal{B}$, the number of lines of the geometry.\\

It can be shown that for the three linear spaces which are the three finite geometries, $\cal{A} $-
affine, $\tilde{\cal{A}}$ - dual affine and $\cal{P}$ - projective geometry
 the number of points on a line $L_j$ i.e. $k_{L_j}$ is independent of the line, j and
 the number of lines sharing a point, $S_{\alpha}$ i.e. $r_{p=S_{\alpha}}$ is independent of
of the point $\alpha$ they are designated by $k_L$ and $r_p$ respectively. The number of points
on a line,  $k_L$, referred to as the "order" of the geometry in \cite{bennett, batten} is dubbed the
"dimensionality" of the geometry
in the present paper since it is what relates to the dimensionality of the underpinned Hilbert space. \\

We now list for each of the geometries four characterizations two of which are sufficient to define the
geometry pertaining to the dimensionality, d, for which the geometry is defined.\\

\noindent $\cal{A}$:     $\nu$ = $d^2$;$\cal{B}$ = d(d+1); $k_L$ = d; $r_p$ = d+1.\\
\noindent $\tilde{\cal{A}}$:      $\nu$ = d(d+1); $\cal{B}$ = $d^2$; $k_L$ = d+1; $r_p$ = d.\\
\noindent $\cal{P}$:        $\nu$ =  $\cal{B}$ = $d^2+d+1$; $k_L$ =  $r_p$ = d+1.\\

Thus, as an example: A linear space with $\nu\;=\;d^2$ and $k_L\;=\;d$ is necessarily an
affine plane geometry, viz, $\cal{A}$  with   $\cal{B}$=d(d+1) and $r_p$=d+1. The proof
is given in an appendix B.\\

We designate a set of parallel (i.e. having no common point) lines as  "pencil".
This is it's mathematical designation \cite{bennett}. (It is referred to as striation by
Wootters  \cite{wootters4}) Thus the structure of the linear space, $\cal{A}$ (i.e. APG)
is accounted by: \textcolor{blue}{The d-dimensional $\cal{A}$ is made of d+1
pencils, each made of d lines. Each line contains d points and has one common point with
every other line of distinct pencil.} Note that the numbers fit:  $\cal{B}$ = d(d+1),
$\nu= d^2$,  $k_L$ = d; $r_p$ = d+1.
For APG,  $r_S=r_p$, $r_S$ being the number of pencils. A proof is given in Appendix C.\\
\textcolor{blue}{In the  dual spaces} pencils are sets of {\it points} with no
common line. Thence  the structure of the dual space $\tilde{\cal{A}}$ is:
\textcolor{blue}
{ A d-dimensional dual linear space $\tilde{\cal{A}}$ is made of d+1 pencils each of
d points not connected by a line. Each point
is common to d lines and has one common line with every point belonging to distinct
pencil. Every two lines
have one point in common, i.e. there are no parallel lines.} For DAPG,
 $\tilde{\cal{A}}$, we have $r_S=k_L$.\\

The remarkable feature characterizing FPP is the dual role played by points and
lines. i.e. interchanging (with the suitable
linguistic adjustment) the words point and line leaves FPP unaffected. Thus
the dual of $\lambda_1,\lambda_2, P1, P2.$ holds for FPP equally well.\\

It can be shown \cite{bennett} that the removal from FPP, of any one line and its points leaves a DAPG.
Removal of
any point and the lines going through it leaves an APG. Conversely given a
APG  upon addition of  a point connecting members of each pencil (set of
parallel lines) and then forming a line with these d+1 points gives a FPP.
And, correspondingly, connecting in DAPG each set of disconnected points with
a line and then having these d+1 lines intersect at a point reproduces FPP.
\textcolor{blue}{Thus a d for which any one of the three geometries exists implies the
existence of the
the other two.} \\

For example, given that APG exists for d=5 - implies the existence, for d=5, of DAPG and
FPP. The characteristics of the APG are, as stipulated above: d=5 $\Rightarrow,\; \nu=25,\;
$\cal{B}$=30,\;r_p=5\;and\;
k_L=6.$ For the DAPG (with d=5) these are $\nu=30,\; $\cal{B}$=25,\;r_p=6 \;and\;k_L=5$.
For FPP, d=5, they are
$\nu= $\cal{B}$=31,\;r_p=k_L=6.$ For $\cal{P}$, i.e. FPP $r_S=0$.\\

No general rule specifying the dimensionalities, d, for which the geometries
exist is known \cite{bennett}.
For $d=p^m$ (power of prime) the geometries may be constructed. No exception to this rule is known.\\

APG is the most intuitive geometry this as  it is closely related to two dimensional vector space in
terms of which it may be coordinated. We now outline its
construction for d=p (a prime).\\

Define a two dimensional vector space ${\cal{V}}: ((x_i,y_j);\;x_i,y_j\in
\mathbb{F}_d\equiv
\mathbb{Z}/d\mathbb{Z},\;i,j=0,1,...d-1)$. i.e. $x_i,y_j$ may be considered as numbers
abiding modular
algebra.  We have, $(x_i;y_j) \in {\cal{V}}$ with addition being  modular addition component wise and
multiplication
by $r\in \mathbb{F}_d$ gives  $(rx_i,ry_j)\in {\cal{V}}$. \\
Now consider a square array of $d^2$ points: d along the "x-axis"  and above each $x_i$
a column of d $y_j$.  The "tip" of a vector, $(x_i,y_j)$, defines a "point". \\
Points (i.e. vectors) whose components  satisfy an equation of the form
\begin{equation}\label{line}
y=rx+s \;Mod[d],\;\;r,s\in \mathbb{F}_d,
\end{equation}
form the line $L_j$, $j=(r,s)$, r,s=0,1,2...d-1. Thus
the equation specifies $d^2$ lines which, with the d lines given by $x=s;\;s=0,1,...d-1,$ gives the d(d+1)
 lines of APG. There are d+1 pencils (striations): the d lines given by the d values of s for each pencil:
 two lines with distinct values of s (holding r fixed or, for the last pencil, holding
${x_i=s}$) gives lines with no common point. If we consider arbitrary pair of lines
belonging to different pencils it is obvious that they share one point: the unique point
$(x_i,y_j)\in {\cal{V}}$ that satisfies the two equations
($r\ne r',\;r,r{'}\ne 0$)
\begin{eqnarray}
y_i&=&rx_j+s,\;Mod[d]\;\;y_i=r{'}x_j+s{'},\;Mod[d]\;\nonumber\\
&\Rightarrow&\;\;x_j=\frac{s{'}-s}{r-r{'}},\;\;y_i=\frac{r{'}r^{-1}s-s{'}}{r{'}r^{-1}-1}.
\end{eqnarray}
(That each line of the vertical pencil, $x=s,$ has a unique common point with each line
of the other pencils is obvious.)\\
Since each point must be common to d+1 lines, one from each pencil, the
construction has $r_p=d+1$. Having the array as square implies $k_L=d$. Thus
we have constructed APG for d=p. The line is parameterized
as, $L_{j=(r,s)}$. \\
The procedure may be described alternatively \cite{dillon} as follows. Define a one
dimensional vectorial subspace, W via: $W:=\{(x_i,rx_i)|r,x_i\in \mathbb{F}_d\}$. The
line, $L_j$ is defined by,
$$L_{j=(r,s)}=(0;s)+W.;\;(o;s)\in{\cal{V}}\;\;s\in \mathbb{F}_d,$$
i.e. a line is a vectorial coset \cite{dillon}.

Forming APG for dimension, d=p (a prime) as was shown above was based on
having, in such cases, consistent modular algebra: The elements (e.g. x,y)
are elements of an (algebraic) field, $\mathbb{F}_d \equiv GF(p))$.\\

As is well known \cite{bennett, schroeder} algebraic fields are also possible
for $d=p^m$, m a positive integer, forming thereby an (algebraic) field
$GF(p^m)$, with the elements $z=0,1,...,p^m-1$ which is an extension of GF(p). An
 illustration of such extension, an extension of  d=3 to $d=3^2$, is given in
Appendix A.\\

Our interest is in "realization" of the geometries, i.e. utilizing {\it geometrical
points}, $S_{\alpha},$ as underpinning Hilbert space states or operators, i.e.
$$S_{\alpha}\Rightarrow \hat{S}_{\alpha}\; or\; |S(\alpha)\rangle,$$
the latter being Hilbert space entities. Now,
given the {\it geometrical} interrelation between lines $L_j$ and points $S_{\alpha}$,
\begin{equation}\label{logical}
L_j\;\equiv\;\bigcup_{\alpha \in j} S_{\alpha},
\end{equation}
 - we seek an implied
interrelation among the  Hilbert space operators (rsp. states) underpinned
with geometrical points - with the corresponding "line" operators,
$\hat{L}_{j}$, (resp. "line states",$|S(\alpha)\rangle$) that relate to those
points via the geometry. (E.g. APG realization as vector field is discussed
above.) Addition is defined for Hilbert space operators and states, we thus
\textcolor{blue}{ define} "line operators" (rsp. "line state"),
$\hat{{\cal{L}}}_j \;(resp.\;|{\cal{S}}(\alpha)\rangle)$ within APG, by (We
use script lettering to emphasize that we deal with APG.)

\begin{equation}\label{def.l}
\hat{{\cal{L}}}_j\;\equiv \;\frac{1}{k_L}\sum_{\alpha \in
j}\hat{{\cal{S}}}_{\alpha},
\end{equation}
to implement the interrelation among points and lines given by
Eq.(\ref{logical}). \textcolor{blue}{Thus the present study considers the
implied Hilbert space {\it "line operators" (resp. "line states"} that follows
from the underpinning of Hilbert space operators (resp. states) with
geometrical points .}\\
Since, as is shown in Section IV, DAPG is more convenient for our study, we
transcribe Eq.(\ref{def.l}) to DAPG:
\begin{equation}\label{real}
\hat{S}_{\alpha}=\frac{1}{d}\sum_{j\in \alpha}\hat{L}_j.
\end{equation}
We used $r_p=d$ within DAPG.

\textcolor{blue}{This definition leads directly to the universal quantity }
\begin{equation}\label{u}
\frac{1}{k_L}\sum_{\alpha}^{d(d+1)}\hat{S}_{\alpha}=\frac{1}{r_p}\sum_j^{d^2}\hat{L}_j
\end{equation}
It is universal in that it involves quantities that are independent of either
lines or points. Here $r_p=d,\;k_L=d+1$.
(The relation holds for APG upon interchanging lines with points.) \\

Proof: (p designates pencil, there are d+1 pencils in a d dimensional DAPG,
cf. Appendix C.)
\begin{eqnarray}
\hat{S}_{\alpha}&=&\frac{1}{r_p}\sum_{j\in \alpha}\hat{L}_j\Rightarrow
\sum_{\alpha\in p}\hat{S}_{\alpha}=\frac{1}{r_p}\sum_{\alpha\in p}\sum_{j\in
\alpha}\hat{L}_j=\frac{1}{r_p}\sum_{j}^{d^2}\hat{L}_j\Rightarrow \nonumber \\
\sum_{p}^{d+1=k_L}\sum_{\alpha\in
p}\hat{S}_{\alpha}&=&\sum_{\alpha}^{d(d+1)}\hat{S}_{\alpha}=\frac{d+1}{d}\sum^{d^2}\hat{L}_j
\Rightarrow \nonumber \\
\frac{1}{d+1}\sum_{\alpha}^{d(d+1)}\hat{S}_{\alpha}&=&\frac{1}{d}\sum^{d^2}_j\hat{L}_j.\;\;\;QED
\end{eqnarray}

This relation reflects the fundamental linear space identity,
 \cite{batten}
\begin{equation}\label{univ}
\sum_{j} d_{L_{j}}=\sum_{\alpha}d_{S_{\alpha}}
\end{equation}
where $d_{L_{j}}$, $d_{S_{\alpha}}$ are, respectively the number of points on
the line j and the number of lines on the point $\alpha$. The
identity is between  two different ways \cite{batten} of summing.\\

The definition, Eq.(\ref{def.l}) within APG, or respectively, Eq.(\ref{real})
within its dual, DAPG, constrains the allowed operators (resp. states) that
may be underpinned with either geometry. Thus let the operators (resp. states)
be separated into ( necessarily) mutually exclusive pencils. Only when the sum
of the distinct members of the distinct  pencils are equal, the operators
(resp. states)
may be underpinned with a geometry. The proof is given in Appendix E.\\

The Hilbert space operators and states considered relates to Mutually Unbiased Bases
that is reviewed below.\\

\section{ Mutually Unbiased Bases (MUB) and Mutually Unbiased Collective Bases (MUCB)}

In a d-dimensional Hilbert space two complete, orthonormal vectorial bases,
${\cal B}_1,\;{\cal B}_2$,
 are said to be MUB if and only if (${\cal B}_1\ne {\cal B}_2)$
\begin{equation}
\forall |u\rangle,\;|v \rangle\; \epsilon \;{\cal B}_1,\;{\cal B}_2
\;resp.,\;\;|\langle u|v\rangle|=1/\sqrt{d}.
\end{equation}
Maximal number of MUB allowed in a d-dimensional Hilbert space is d+1
\cite{ivanovich,bengtsson}. Variety of methods for construction of the d+1
bases for $d=p^m$ are now available
\cite{wootters2,tal,klimov2,vourdas}. Our present study is confined to $d=p\;(a\; prime)\;\ne 2$.
We comment on the cases $d=p^m,\;m>1$ in the Appendix. In such cases the modular variables, n,b etc.,
are elements of an (algebraic) field $GF(p^m)$. \\
It is convenient \cite{wootters2,tal,klimov2,vourdas} to list the d+1 MUB
bases in terms of the so called computational basis (CB). The CB states
$|n\rangle,\;n=0,1,..d-1,\;|n+d\rangle=|n\rangle,$
 are eigenfunction of $\hat{Z}$,
\begin{equation}\label{z}
\hat{Z}|n\rangle=\omega^{n}|n\rangle;\;\omega=e^{i2\pi/d},
\end{equation}
We now give explicitly the MUB states in conjunction with the algebraically
complete operators \cite{schwinger,amir} set, $\hat{Z},$ and the shift
operator, $\hat{X}|n\rangle=|n+1\rangle:$ In addition to the CB the d other
bases, each labeled by b, are \cite{tal}
\begin{equation} \label{mxel}
|m;b\rangle=\frac{1}{\sqrt
d}\sum_0^{d-1}\omega^{\frac{b}{2}n(n-1)-nm}|n\rangle;\;\;b,m=0,1,..d-1,
\end{equation}
the m labels states within a basis. Each basis relates to a unitary operator,
\cite{tal}, $\hat{X}\hat{Z}^b|m;b\rangle=\omega^m|m;b\rangle.$ For later
reference we shall refer to the computational basis (CB) by $b=\ddot{0}$. Thus
the  d+1 bases, are, $b=\ddot{0}$ and b=0,1,...d-1. The total number of states
is d(d+1) they are grouped in d+1 sets each of d states. When no confusion may
arise we abbreviate the states in the CB $|m,\ddot{0}\rangle,$ i.e. the state
m in the basis $\ddot{0}$, by $|\ddot{m}\rangle,$ or simply $|m\rangle;$ we
abbreviate
$|m,0\rangle$, i.e. the m state in the basis b=0 by $|m_0\rangle$.\\
We choose the phase of the CB nil, and note  that the MUB  set is closed under
complex conjugation,
\begin{eqnarray}\label{cc}
\langle n|m,b\rangle^{\ast}&=&\langle
n|\tilde{m},\tilde{b}\rangle,\;\Rightarrow|\tilde{m},\tilde{b}\rangle
=|d-m,d-b\rangle,\;b\ne\ddot{0},\nonumber \\
\langle n|m\rangle&=&\langle n|m\rangle^{\ast},\;b=\ddot{0}.
\end{eqnarray}
as can be verified from Eq.(\ref{mxel}).\\
Several studies \cite{durt,klimov1,berge,rev1} consider the entanglement
of two d-dimensional particles Hilbert space via MUB state labeling. We shall
now outline briefly
the approach adopted by \cite{rev1} that will be used in later sections.\\

Guided by the continuum case, $d\rightarrow \infty$ \cite{rev1}, where it is
natural to consider collective coordinates (and operators) that refer to
relative and center of mass coordinates we consider the definitions for
"relative" and "center of mass" for the finite dimensional Hilbert spaces.
Thus the Hilbert space is spanned by the single particle computational bases,
$|n\rangle_1|n'\rangle_2$ (the subscripts denote the particles). These are
eigenfunctions of $\hat{Z}_i$ i=1,2:
$\hat{Z}_i|n\rangle_i=\omega^{n}|n\rangle_i,\;\omega=e^{i\frac{2\pi}{d}}.$
Similarly $\hat{X}_i|n\rangle_i=|n+1\rangle_i,\;i=1,2$. We now define our
collective coordinates and collective operators (we remind the reader that the
exponents are modular variables, e.g. 1/2 mod[d=7]=(d+1)/2)=4):
\begin{equation}\label{colz}
\hat{Z}_r\equiv \hat{Z}^{1/2}_{1}\hat{Z}^{-1/2}_{2};\;\;\bar{Z}_c\equiv
\hat{Z}^{1/2}_{1}\hat{Z}^{1/2}_{2}\;\leftrightarrow\;\hat{Z}_1=\hat{Z}_r\hat{Z}_c;\;\;\hat{Z}_2=
\hat{Z}_r^{-1}\hat{Z}_c,
\end{equation}
and, in a similar manner,
\begin{equation}\label{colx}
\hat{X}_r\equiv\hat{X}_1\hat{X}_2^{-1};\;\hat{X}_c\equiv\hat{X}_1\hat{X}_2\leftrightarrow
\hat{X}_1=\hat{X}^{1/2}_r\hat{X}^{1/2}_c,\;\hat{X}_2=\hat{X}^{-1/2}_r\hat{X}^{1/2}_c.
\end{equation}

We note that $\hat{Z}_{s}^{d}=\hat{X}_s^{d}=1,$  and
$\hat{X}_s\hat{Z}_s=\omega\hat{Z}_s\hat{X}_s,\;s=r,c;
\;\hat{X}_s\hat{Z}_{s'}=\hat{Z}_{s'}\hat{X}_s,\;s\ne s'.$
$|n_1\rangle|n_2\rangle,$ the eigenfunctions of $\hat{Z}_i,\;i=1,2,$
 spans the $d^2$ dimensional Hilbert space.  The sets $\hat{Z}_i,\hat{X}_i;\;i=1,2$ are
 algebraically complete in this space \cite{schwinger}, i.e. every (non trivial) operator
 is a function of these operators. The eigenfunctions of $\hat{Z}_q$ are $|n_c,n_r\rangle$ with
$\hat{Z}_c|n_c,n_r\rangle=\omega^{n_c}|n_c,n_r\rangle,\;\hat{Z}_r|n_c,n_r\rangle=
\omega^{n_r}|n_c,n_r\rangle.$ We note, e.g. \cite{sch}, that $|n_c,n_r\rangle$
is equivalent to $|n_c\rangle|n_r\rangle$ when, as is the present case, the
two sets,
$\hat{Z}_q,\hat{X}_q;\;q=c,r$ are compatible.\\

Clearly $|n\rangle_r|n'\rangle_c;\;n,n'=0,1,..d-1,$ is a $d^2$ orthonormal
basis spanning the two d-dimensional particles Hilbert space. We may consider
their respective computational eigen-bases and with them the whole set of MUB
bases \cite{rev1},
\begin{equation}
\hat{Z}_{s}|n\rangle_s = \omega^{n}|n\rangle_s,\;\;
\hat{X}_s\hat{Z}_s^{b_s}|m_s,b_s\rangle=
\omega^{m_s}|m_s,b_s\rangle;\;\;\langle
n_s|m_s,b_s\rangle=\omega^{\frac{b_s}{2}n_s(n_s-1)-m_sn_s}.\;s=r,c.
\end{equation}

States in the particle coordinates may, clearly, be expressed in terms of the
product states of the collective coordinates as both form a complete
orthonormal basis that span the two particles d-dimensional Hilbert space,
\begin{equation}
|n_1\rangle|n_2\rangle=\sum_{n_c,n_r}|n_c,n_r\rangle\langle n_c,
n_r|n_1\rangle|n_2\rangle.
\end{equation}
The matrix element $\langle n_c, n_r|n_1\rangle|n_2\rangle$ is readily
evaluated \cite{rev1},

\begin{equation}\label{relcomdel}
\langle
n_1,n_2|n_r,n_c\rangle=\delta_{n_r,(n_1-n_2)/2}\delta_{n_c,(n_1+n_2)/2}.
\end{equation}
We have then,
\begin{equation}\label{relcom}
|n_r,n_c\rangle
\Leftrightarrow|n_1,n_2\rangle,\;\;for\;n_r=(n_1-n_2)/2,\;n_c=(n_1+n_2)/2\;\rightleftarrows
n_1=n_r+n_c,\;n_2=n_c-n_r.
\end{equation}
 There are,
of course, d+1 MUB bases for each of the collective modes. Here too, we adopt
the notational simplification $b_s\rightarrow \ddot{0}_s,\; s=r,c$.

\section{ Underpinning Mutual Unbiased Bases (MUB) with Dual Affine Plane Geometry (DAPG)}

For d=prime consider an array of d(d+1) points arranged as d+1 columns of d points each. We specify
each point by two indices (b,m). b specifies the column: b=$\ddot{0}$ designated the left - most column,
the column next to it is b=0, the next b=1 and so on with the right-most column being b= d-1.\\
The rows are specified by m. m=0 is the upper most row, the row below it is dubbed m=1 and so on. The
bottom
row is m=d-1.\\
We now consider the points as underpinning MUB projector. Recalling, Eq.(\ref{u}), that for d=prime

$$|m;b\rangle=\frac{1}{\sqrt d}\sum_{0}^{d-1}\omega^{\frac{b}{2}n(n-1)-mn}|n\rangle;\;\;b=0,1,...d-1;\;\;
|m;\ddot{0}\rangle=|m\rangle,\;\;\;(10a)$$
the MUB projector is given by
\begin{equation}\label{proj}
\mathbb{P}(m,b)=|m;b\rangle\langle b;m|.
\end{equation}
Thus, e.g., for d=3 the array is
\begin{equation}\label{pr}
 \left( \begin{array}{ccccc}
m\backslash b&\ddot{0}&0&1&2 \\
0&\mathbb{P}(0,\ddot{0})&\mathbb{P}(0,0)&\mathbb{P}(0,1)&\mathbb{P}(0,2)\\
1&\mathbb{P}(1,\ddot{0})&\mathbb{P}(1,0)&\mathbb{P}(1,1)&\mathbb{P}(1,2)\\
2&\mathbb{P}(2,\ddot{0})&\mathbb{P}(2,0)&\mathbb{P}(2,1)&\mathbb{P}(2,2)\end{array} \right).\\
\end{equation}

We use the d(d+1) DAPG points to underpin the d(d+1) MUB projectors. A line
contain one point from each column. (Ensuring $k_L=d+1.$) The d+1 DAPG pencils
(i.e. set of points not connected by a line) underpin the d+1 MUB bases. The d
member of each pencil underpin the d orthogonal
projectors of a basis b (cf. Appendix C).\\

We now derive the line equations in two steps. First we obtain the equation
for part of the line, the part containing one point from each column  b=0 to
b=d-1. We refer to this as an amputated line (AL), and designate it by L'. It
has d points. Later we obtain the (d+1)st point dubbed L" that with L form to
full line, L. We require the CB representation of the MUB projectors.\\

a. Derivation of the amputated line (AL) , L', equation.\\

 Using  Eq.(\ref{mxel}) and Eq.(\ref{proj}) for b=0,1,...d-1
\begin{equation}\label{mxmub}
\langle n|\mathbb{P}(m,b)|n{'}\rangle=\frac{1}{d}\omega^{i(n-n{'})[\frac{b}{2}(n+n{'}-1)-m]};\;
b=0,1,...d-1.
\end{equation}
The followings are implied (trivially) by Eq.(\ref{mxmub}):
\begin{eqnarray}\label{li}
\langle n|m;b\rangle\langle b;m|n{'}\rangle&=&\langle n|m{'};b{'}\rangle\langle b{'};m{'}|n{'}\rangle;\;\;
b\ne b{'}\nonumber \\
\Rightarrow&&\frac{b}{2}(n+n{'}-1)-m\;=\; \frac{b{'}}{2}(n+n{'}-1)-m{'}\nonumber \\
\Rightarrow&& m(b)=m_0+\frac{b}{2}(n+n{'}-1);\;m_0=m(0). \nonumber \\
\langle s|m;b\rangle\langle b;m|s{'}\rangle&=&\langle s|m{'};b{'}\rangle\langle b{'};m{'}|s{'}\rangle;\;\;
b\ne b{'};\;s\ne s{'}.\nonumber \\
\Leftrightarrow&&s+s{'}=n+n{'}. \nonumber \\
\langle n|\bar{m};b\rangle\langle b;\bar{m}|n{'}\rangle&=&\langle n|\bar{m}{'};b{'}\rangle\langle b{'};
\bar{m}{'}|n{'}\rangle;\;\;\bar{m}\ne m;\nonumber \\
\Rightarrow&&\bar{m}(b)=\bar{m}_0+\frac{b}{2}(n+n{'}-1);\;\bar{m}_0=\bar{m}(0)\nonumber\\
\bar{m}\ne m &\Rightarrow&m_0\ne\bar{m}_0.
\end{eqnarray}

Requiring equality of the n,n' matrix elements of the projectors in distinct
columns implies a "line equation", viz the value of m (row) as a function of b
(column) for which the equality holds. Via Eq.(\ref{li}) this is shown  to
yield
$$m(b)=m_0+bc-b/2;\;\;m_0\equiv m(0),\;\;2c=n+n{'};\;\;b=0,1,2...d-1.$$
In Eq.(\ref{li}) it is also shown that the equality of the n,n' matrix
elements implies the equality of the s,s' matrix elements provided
$$s+s{'}=n+n{'}\equiv 2c.$$
Thus selecting arbitrary m in arbitrary b, $b\ne \ddot{0},$ and requiring the
equality of the n,n' matrix elements of in all the columns, b ($b\ne
\ddot{0}$) determines AL equation m(b) as stipulated above, dubbed
$L{'}_{m_0,c}$. i.e. it is parameterized by $m_0,c$. Selecting a different
$\bar{m}\ne m$ and imposing the same requirements gives a different AL,
$L{'}_{\bar{m}_0;c}$. \textcolor{blue}{Thus distinct amputated lines (AL)
parameterized with the same value of c, have no common point}:

$$\bar{m}(b)=\bar{m}_0+bc-b/2;\;\;\bar{m}_0\equiv \bar{m}(0),\;\;2c=n+n{'};\;\;b=0,1,2...d-1.$$
The barred AL, $L{'}_{\bar{m}_0,c},$ has no common point with the unbarred
one, $L{'}_{m_0,c}$.

Since every pair of full DAPG  lines do share one point, the full lines
extension of all the amputated lines parameterized by c (e.g.  $L{'}_{m_0,c}$
and $L{'}_{\bar{m}_0,c}$) must share the point at $b=\ddot{0}$. I.e.
$n+n{'}\Rightarrow \ddot{m}.$ The "natural" relation is
$$n+n{'}\;\equiv 2c =\;2\ddot{m}.$$ We adopt this relation.(Alternatives to this
are considered at the end of this section, in subsection d.)
 The full line equation is, then,
\begin{eqnarray}\label{eql}
m(b)&=&m_0+b\ddot{m}-b/2;\;\;b=0,1,...d-1. \nonumber \\
    &=& \ddot{m};\;\;b=\ddot{0}.
\end{eqnarray}
The upper part pertains to the AL. We may thus parameterize both the AL and
the (full) line with $\ddot{m},m_0:$ $L'(\ddot{m},m_0)$ and $L_{\ddot{m},m_0}$
respectively. It is now obvious that  every two (full) lines parameterized by
$\ddot{m}$ (or equivalently with c) have one common point - indeed it is
$\ddot{m}$ in the $b=\ddot{0}$ column.\\
The complete proof of the $c\Leftrightarrow \ddot{m}$ relation obtains upon
noting that for $c \ne c{'}$ the two AL do have a common point at
$b\ne\ddot{0}$. Thus,
$$m_0+bc=\bar{m}_0+bc{'}\;\Rightarrow b=(m_0-\bar{m}_0)/(c{'}-c),$$
giving for $m_0 \ne\bar{m}_0$ gives as the common point $m=m_0+bc-b/2$ at
$b=(m_0-\bar{m}_0)/(c{'}-c)$. For $m_0=\bar{m}_0$ the common point is $m=m_0$
at $b=0$. Thus \textcolor{blue}{two amputated lines parameterized with distinct values of
c do have a common point.} Thus there is a 1-1 relation between c and $\ddot{m}$\\

We pause here to verify that our underpinning arena, viz  the d(d+1) points
$S_{\alpha};\alpha=(m;b)$ is indeed a DAPG. The points array specified by d+1
columns labeled with b, $b=\ddot{0},0,1,...d-1$, (corresponding to MUB bases)
and d rows labeled with m, m=0,1,...d-1, (corresponding to MUB vectors within
a basis). The $d^2$ lines are specified via, Eq.(\ref{eql}) (corresponding to
$\ddot{m},m_0=0,1,2,...d-1$.)\\

Hence,\\

\noindent 1. $\nu=d(d+1);$\; ${\cal{B}}=d^2.$\\
\noindent 2. $k_L=d+1.$\\
\noindent3. There are d+1 sets of points, each containing d points that have no interconnecting lines, i.e.
   d+1 pencils. (These are the d points in each column.)\\
\noindent 4. Every line has one common point with every other line.\\
\noindent 5. $r_p=d.$\\
\noindent Items 1., 2. and 3. are obvious. Item 4. may be seen from
Eq.(\ref{eql}): two line equations allow
one and only one solution for a common m value. Item 5. follows from 3. and 4. Thus the lines and
points considered form a DAPG.\\

Thus DAPG forms a natural underpinning array for the projectors of MUB, as is illustrated in Eq.(\ref{pr})
above. Both the geometry and MUB may be constructed for d=p (prime) that is considered here.
The point specified by (m,b) underpins the projector $\mathbb{P}(m,b)$.\\

b. Explicit form of the AL operator, $\hat{L}'_{\ddot{m},m_0}$, and the (full) line operator,
$\hat{L}_{\ddot{m},m_0}$.\\

 We now derive the Hilbert  space operator $\hat{L}(\ddot{m},m_0)$, ( $\ddot{m}\equiv m(\ddot{0}),
 m_0\equiv m(0)$ define the line) that is underpinned with the geometrical (DAPG) line formed by the
 points, $\hat{S}_{(\alpha=(m,b))}$. These constitutes the line specified by $\ddot{m},m_0$ via
 Eq.(\ref{eql}).\\
Utilizing our formulas derived in Section II we extract, quite generally, within DAPG, for arbitrary d
(=prime)
the expression for $\hat{L}_{j}$ in terms of $\hat{S}_{\alpha}$ starting with the definition,
Eq.(\ref{def.l}).
\begin{eqnarray}\label{lop}
\hat{S}_{(\alpha=(m,b))}&=&\frac{1}{r_p}\sum_{j\in\alpha} \hat{L}_j \Rightarrow
\sum_{\alpha\in j}\hat{S}_{(\alpha=(m,b))}=\frac{1}{d}\sum_{\alpha\in j}\sum_{j{'}\in \alpha}\hat{L}_{j{'}}
= \nonumber \\
&=&\frac{1}{d}\big[(d+1)\hat{L}_j+\sum_{j{'}\ne j}\hat{L}_{j{'}}\big]=
\hat{L}_j + \frac{1}{d}\sum_{j{'}}^{d^2}\hat{L}_{j{'}},\Rightarrow \nonumber \\
\hat{L}_j=\sum_{\alpha\in j\ni}\hat{S}_{\alpha} -\frac{1}{d+1}\sum_{\alpha}\hat{S}&=&
\sum_{\alpha\in j}\hat{S}_{\alpha}-\mathbb{I}\;=\;\sum_{b=\ddot{0}}^{d-1}\mathbb{P}(m(b);b)-\mathbb{I}.
\end{eqnarray}
Where we used the universal relation, Eq(\ref{u}) that gives a summation over
all the points, i.e. over all MUB projectors. Summing over the MUB projectors
gives d+1 times unity $\mathbb{I}$, e.g. consider Eq.(\ref{pr}) for d=3:
summing over the projectors in each of the 4 (=d+1) columns gives $\mathbb{I}$.\\
We evaluate the AL contribution first,
\begin{equation}
\hat{L}'(\ddot{m};m_0)=\sum_{b=0}^{d-1}\mathbb{P}(m(b);b)-\mathbb{I}.
\end{equation}
Since the diagonal elements of the d projectors are 1/d subtracting
$\mathbb{I}$ leaves
$$\langle n|\hat{L}'(\ddot{m};m_0)|n\rangle = 0,\;\;n=0,1,...d-1.$$
Since, via our definition of the line, Eq.(\ref{li}) the n,n{'} matrix
elements of all the line elements are equal
$$\langle n|\mathbb{P}(m(b);b)|n{'}\rangle=\frac{\omega^{-(n-n{'})m_0}}{d}\;\;\forall n,n{'}
\;such \;that \;n+n{'}=2\ddot{m}.$$ Thence, the non vanishing matrix elements
for AL are,
$$\langle n|\hat{L}'(\ddot{m};m_0)|n{'}\rangle=\omega^{-(n-n{'})m_0};\;\;n+n{'}=2\ddot{m}.$$
Since for n,n{'} with $n+n{'}\ne 2\ddot{m}$ no two terms are equal, i.e.
$$\langle r|\mathbb{P}(m(b);b)|r{'}\rangle\ne\langle r|\mathbb{P}(m(b{'});b{'})|r{'}\rangle;\;\;b\ne b{'}\;
and\;r+r{'}\ne 2\ddot{m}.$$
Thence the sum over b from b=0 to b=d-1 sums over the d roots of unity hence
$$\sum_{b=0}^{d-1}\langle r|\mathbb{P}(m(b);b)|r{'}\rangle=0\;\;r+r{'}\ne 2\ddot{m}.$$
Thus,
\begin{eqnarray}
\langle n|\hat{L}'(\ddot{m};m_0)|n{'}\rangle&=&\begin{cases} \delta_{n+n{'},2\ddot{m}}\omega^{-(n-n{'})m_0},\;
n\ne n{'} \\0,\;n=n{'}.\end{cases} \nonumber \\
\Rightarrow\langle n|\hat{L}_{\ddot{m};m_0}|n{'}\rangle&=&\delta_{n+n{'},2\ddot{m}}\omega^{-(n-n{'})m_0}.
\end{eqnarray}

This is  illustrated now for the d=3. We first give the projectors in the CB representation, cf.
Eq.(\ref{mxel},10a,\ref{pr})

\begin{eqnarray}
\mathbb{P}(0,\ddot{0})&=&\begin{pmatrix}1&0&0\\0&0&0\\0&0&0\end{pmatrix};\;
\mathbb{P}(0,0)=\;\frac{1}{3}\begin{pmatrix}1&1&1\\1&1&1\\1&1&1\end{pmatrix};\;\;\;\;
\mathbb{P}(0,1)=\frac{1}{3}\begin{pmatrix}1&1&\omega\\1&1&\omega\\\omega^2&\omega^2&1\end{pmatrix};\;
\mathbb{P}(0,2)=\frac{1}{3}\begin{pmatrix}1&1&\omega^2\\1&1&\omega^2\\\omega&\omega&1\end{pmatrix};\nonumber \\
\mathbb{P}(1,\ddot{0})&=&\begin{pmatrix}0&0&0\\0&1&0\\0&0&0\end{pmatrix};\;
\mathbb{P}(1,0)=\frac{1}{3}\begin{pmatrix}1&\omega^2&\omega\\\omega&1&\omega^2\\\omega^2&\omega&1\end{pmatrix};\;
\mathbb{P}(1,1)=\frac{1}{3}\begin{pmatrix}1&\omega^2&\omega^2\\\omega&1&1\\\omega&1&1\end{pmatrix};\;
\mathbb{P}(1,2)=\frac{1}{3}\begin{pmatrix}1&\omega^2&1\\\omega&1&\omega\\1&\omega^2&1\end{pmatrix};\nonumber \\
\mathbb{P}(2,\ddot{0})&=&\begin{pmatrix}0&0&0\\0&0&0\\0&0&1\end{pmatrix};\;
\mathbb{P}(2,0)=\frac{1}{3}\begin{pmatrix}1&\omega&\omega\\\omega^2&1&\omega\\\omega&\omega^2&1\end{pmatrix};\;
\mathbb{P}(2,1)=\frac{1}{3}\begin{pmatrix}1&\omega&1\\\omega^2&1&\omega^2\\1&\omega&1\end{pmatrix};\;
\mathbb{P}(2,2)=\frac{1}{3}\begin{pmatrix}1&\omega&\omega\\\omega^2&1&1\\\omega^2&1&1\end{pmatrix};
\end{eqnarray}

We select n=0,n'=2 for m=1, b=0 , i.e. for $\mathbb{P}(1,0)$: $\langle 0|\mathbb{P}(1,0)|2\rangle=
\frac{\omega}{3}$. In b=1 we find that $\langle 0|\mathbb{P}(0,1)|2\rangle=\frac{\omega}{3}$,
i.e. in the column b=1, m=0 gives the same matrix element. For b=2 $\langle 0|\mathbb{P}(2,2)|2\rangle=
\frac{\omega}{3}$. One notes that these three matrices have equal matrix elements for n=2, n'=0 i.e. with
the same n+n':
$$\langle 0|\mathbb{P}(1,0)|2\rangle=\langle 0|\mathbb{P}(0,1)|2\rangle=\langle 0|\mathbb{P}(2,2)|2\rangle
=\frac{\omega^2}{3}$$
The projector for $b=\ddot{0}$ is $\mathbb{P}(1,\ddot{0})$ since $n+n'=2\rightarrow 2m=2\rightarrow m=1$.
Thus the line, viz m(b) is:
$$m(\ddot{0})=1,\;m(0)=1,\;m(1)=0,\;m(2)=2.$$
The points forming this line are marked with $*$,
\[ \left( \begin{array}{ccccc}
m\backslash b&\ddot{0}&0&1&2 \\
0&-&-&*&-\\
1&*&*&-&-\\
2&-&-&-&*\end{array} \right)\].\\
Evaluating the line operator,
\begin{eqnarray}
\hat{L}_{\ddot{m}=1;m_0=1}&=&\mathbb{P}(1;\ddot{0})+\mathbb{P}(1;0)+\mathbb{P}(0;1)+\mathbb{P}(2;2)-\mathbb{I}
\nonumber \\
&=&\begin{pmatrix}0&0&0\\0&1&0\\0&0&0\end{pmatrix}+
\frac{1}{3}\begin{pmatrix}1&\omega^2&\omega\\\omega&1&\omega^2\\\omega^2&\omega&1\end{pmatrix}+
\frac{1}{3}\begin{pmatrix}1&1&\omega\\1&1&\omega\\\omega^2&\omega^2&1\end{pmatrix}+
\frac{1}{3}\begin{pmatrix}1&\omega&\omega\\\omega^2&1&1\\\omega^2&1&1\end{pmatrix}-
\begin{pmatrix}1&0&0\\0&1&0\\0&0&1\end{pmatrix}=\begin{pmatrix}0&0&\omega\\0&1&0\\\omega^2&0&0\end{pmatrix}.
\end{eqnarray}

The line operators, $\hat{L}_{\ddot{m},m_0}$, are mutually orthogonal,
\begin{eqnarray}
tr \hat{L}_{\ddot{m},m_0}\hat{L}_{\ddot{m{'}},m{'}_0}&=&\sum_{n,n{'}}\langle n|\hat{L}_{\ddot{m},m_0}|n{'}\rangle\langle n{'}|\hat{L}_{\ddot{m},m_0}|n\rangle= \nonumber \\
&=&\sum_{n,n{'}}\delta_{n+n{'},2\ddot{m}}\omega^{-(n-n{'})m_0}\delta_{n+n{'},2\ddot{m{'}}}\omega^{-(n{'}-n)m_0}
=d\delta_{\ddot{m},\ddot{m{'}}}\delta_{m_0,m{'}_0}.
\end{eqnarray}
\textcolor{blue}{Thus an arbitrary Hilbert space operator, $\hat{A}$, is expressible in terms of the
"line operators",
\begin{equation}
\hat{A}=\frac{1}{d}\sum_{j=0}^{d^2-1} \big(tr\hat{A}\hat{L}_j\big)\hat{L}_j,\;\;j=(\ddot{m},m_0),
\end{equation}
in the sense that, for arbitrary two operators $\hat{A}, \;\hat{B},$
\begin{equation}
tr \hat{A}\hat{B}=\frac{1}{d}\sum_{j} tr\hat{A}\hat{L}_j\;tr\hat{B}\hat{L}_j.
\end{equation}
i.e. we have a finite dimensional phase space (i.e. via c-number functions of $(\ddot{m},m_0)$) map of
Hilbert space.}\\

We now argue that the "line operator"  $\hat{L}_{j=(\ddot{m},m_0)}$ is displaced parity operator \cite{grossman,royer,bishop},
Thus
\begin{equation}
\langle n|\hat{L}_{(0,0)}|n{'}\rangle =\delta_{n+n,0}\Rightarrow\hat{L}_{(0,0)}=
\sum_{s=0}^{d-1}|s\rangle\langle -s|\equiv {\cal{I}}.
\end{equation}
Thus "symmetric line operator", i.e. line operator underpinned with line
defined by $c=\ddot{m}$, is \textcolor{blue}{ is a displaced parity operator,}
\begin{eqnarray}\label{shiftinversion}
\hat{L}_{j=(\ddot{m},m_0)}&=&\hat{X}^{\ddot{m}}\hat{Z}^{-m_0}\hat{L}_{(0,0)}\hat{Z}^{m_0}\hat{X}^{-\ddot{m}}
\;\;\Rightarrow \nonumber \\
\hat{L}^2_{(\ddot{m},m_0)}&=&\mathbb{I}.
\end{eqnarray}
Proof of this is given in Appendix D. It expresses the geometrical origin of
the displaced parity operator \cite{grossman,royer,bishop} present in c-number
functions that emulate Hilbert space operators in phase space formulation of
quantum mechanics. The formula agrees with Wootters \cite{wootters4} where it
is given as a point in what is essentially APG.\\
The essential equivalence of both geometries,
APG and DAPG is explained in the succeeding section.\\

c. Affine Plane Geometry (APG) and Dual Affine Plane Geometry (DAPG).\\

We now outline the reasoning that allow viewing \textcolor{blue}{ APG} as closely reminiscent of classical phase space. Our starting point is to consider $(\ddot{m},m_0)$ coordinate of a point in
 (a finite dimensional)
phase space: $\ddot{m},$ CB eigenvalue, playing the role of q and $m_0$, Fourier transform of
the CB, playing the role of p. The $\ddot{m}$ axis is along the  horizontal (i.e. "x axis") and $m_0$ along the vertical, ("y axis"). Within this intuitive language
\textcolor{blue}{ APG  points} underpin $\hat{L}_{\ddot{m},m_0}$,  Hilbert space operator (termed "line operator"
within the DAPG). Similarly \textcolor{blue}{lines} are now designated by $S_{\alpha} (\;\alpha=(m,b))$ (they were designated  points in our DAPG considerations). Correspondingly the expression for the \textcolor{blue}
{line} operator within APG is given by
\begin{equation}\label{apgline}
\hat{S}_{\alpha}=\frac{1}{d}\sum_{j\in \alpha}\hat{L}_j;\;\;j\equiv (\ddot{m},m_0),\;\alpha\equiv (m,b).
\end{equation}
This equation reads: The APG \textcolor{blue}{"line operator", $\hat{S}_{\alpha}$} is the (normalized) sum of the
$\hat{L}_{j}$ that form the line $\alpha$, i.e. that belong to the line equation. We now obtain the actual value of $\alpha$ implied by the line equation.\\ 
We consider phase space points with linear relations between $m_0$ and  $\ddot{m}$ 
emulating thereby "straight lines" that were the starting point in \cite{wootters4, klimov}. We shall show that 
these will give for the LHS, viz. $\hat{S}_{\alpha}$, an MUB projector and relate the line parameters to the (r,s,s') appearing in the equation below to $\alpha$. Thus we consider the following d(d+1) lines,
\begin{eqnarray}\label{dlines}
m_0=r\ddot{m}+s,\;Mod[d]\;\;r,s=0,1,...d-1, \;\;d^2 \;lines.\nonumber \\
\ddot{m}=s',\;Mod[d]\;s'=0,1,...d-1\;\; d\;lines.
\end{eqnarray}

We deal first with the set of $d^2$ lines. The "line operator" \textcolor{blue}{within APG} is given by 
Eq.(\ref{apgline} ).\\ 

The sum involves matrices each with its distinct skew line of non vanishing matrix elements. Their sum
gives the MUB projector, i.e.

\begin{equation}\label{apg=dapg}
 \hat{S}_{(r,s)}=\mathbb{P}(m=s-b/2;b=-r)\equiv|m=s-b/2;b=-r\rangle\langle b=-r;m=s-b/2|.
\end{equation}

 The proof follows from the following reasoning. The $n,n{'}$ matrix element of the MUB projector, Eq.() is
$$\langle n|\mathbb{P}(m;b)|n{'}\rangle =\frac{1}{d}\omega^{(n-n{'})(b/2[n+n{'}-1]-m)}.$$
For these $n,n{'}$ the matrix elements of the APG line operator, labeled with
r and s are, \textcolor{blue}{for $n+n{'}=2\ddot{m}$ and $m_0=r\ddot{m}+s.$}
\begin{equation}
\langle n|\hat{S}_{(r,s)}|n{'}\rangle=\frac{1}{d}\omega^{-(n-n{'})(r\ddot{m}+s)}.
\end{equation}
The two expressions are equal for $r=-b,\;and \;m=s-b/2$. QED.\\
We now consider the set of d lines given by Eq.(\ref{dlines}). In these cases the MUB projector is
$$\mathbb{P}(m=s';b=\ddot{0})=|s'\rangle\langle s'|.$$
The proof is as follows.
\begin{equation}
\langle n|\hat{S}_{s'}|n{'}\rangle=\frac{1}{d}\sum_{m_0=0}^{d-1}\langle n|\hat{L}_{s';m_0}|n{'}\rangle=
\frac{1}{d}\sum_{m_0=0}^{d-1}\delta_{n+n{'},2s}\omega^{-(n-n{'})m_0}.
\end{equation}
The non diagonal terms add up to nil. The diagonal, i.e. $n=n{'}=s'$, add up to 1. QED.

 We illustrate this for d=3 and r=1,s=0:
The APG line is made of the following three points $(\ddot{m}=0;m_0=0)$;$(\ddot{m}=1;m_0=1)$;$\ddot{m}=2;
m_0=2).$
These underpin the  "points operators", given in terms of matrix elements,
\begin{equation}
(\ddot{m}=0;m_0=0)\Rightarrow\delta_{n+n{'},0};\;;(\ddot{m}=1;m_0=1)\Rightarrow\delta_{n+n{'},2}
\omega^{-(n-n{'})};
\;;(\ddot{m}=2;m_0=2)\Rightarrow \delta_{n+n{'},1}\omega^{-(n-n{'})2}.
\end{equation}
Thence, for r=1 and s=0,
\begin{equation}
\hat{S}_{(r=1,s=3)}=\frac{1}{3}\begin{pmatrix}1&0&0\\0&0&1\\0&1&0\end{pmatrix}+
\frac{1}{3}\begin{pmatrix}0&0&\omega\\0&1&0\\\omega^2&0&0\end{pmatrix}+
\frac{1}{3}\begin{pmatrix}0&\omega&0\\\omega^2&0&0\\0&0&1\end{pmatrix}=
\frac{1}{3}\begin{pmatrix}1&\omega&\omega\\\omega^2&1&1\\\omega^2&1&1\end{pmatrix}=\mathbb{P}(2,2).
\end{equation}
i.e. b=-r=-1=2 Mod[3], and m=s-b/2=-b/2=2\;Mod[3].\\
For the case of vertical, i.e. $\ddot{m}=s$, lines we have in the APG the three points
$$(\ddot{m}=s,m_0=0);(\ddot{m}=s;m_0=1);(\ddot{m}=s;m_0=2).$$ In terms of the full matrices this is for s=0:
\begin{equation}
\hat{S}_{r=1,s=3}=\frac{1}{3}\begin{pmatrix}1&0&0\\0&0&1\\0&1&0\end{pmatrix}+
\frac{1}{3}\begin{pmatrix}1&0&0\\0&0&\omega^2\\0&\omega&0\end{pmatrix}+
\frac{1}{3}\begin{pmatrix}1&0&0\\0&0&\omega\\0&\omega^2&0\end{pmatrix}=
\begin{pmatrix}1&0&0\\0&0&0\\0&0&0\end{pmatrix}=\mathbb{P}(0,\ddot{0})
=|0\rangle\langle 0|.
\end{equation}

d. Alternative phase space mapping.\\

We now consider alternative schemes for adjoining L", a point in $b=\ddot{0}$
column, to L', the AL,  to that used in the "symmetric" scheme, viz
$c=\ddot{m}$, considered above. We now consider a
 general linear relation between c and $\ddot{m}$. This is expressed  in the following  line equations,
 $r,s \in \mathbb{F}_d,\;r\ne 0$:\\
\begin{equation}\label{lambda}
L^{(r,s)}_{\ddot{m},m_0}:\;\;m(b)=\begin{cases}m_0+bc-b/2,\;\;b\ne \ddot{0};\\
m(\ddot{0})\equiv \ddot{m}=rc+s, \;\;b=\ddot{0}.\end{cases}
\end{equation}

Each of the line equations accounts for $d^2$ lines, $\ddot{m},m_0=0,1,...d-1$
for fixed r,s ($r\ne 0$). Thus (r,s) defines a family of of $d^2$ lines.  For
r=1 and s=0 the line equation reduces to the "symmetric" one. Note that the
parametrization of
$L^t_{\ddot{m},m_0};\;t=(r,s)$ are, aside from their classification (r,s), like those of the
"symmetric" lines: $L_{\ddot{m},m_0}$, viz $\ddot{m};m_0$. \\

We now prove that the  line operators  $\hat{L}^t_{\ddot{m},m_0},\;t=(s,r)$ form
a set of $d^2$ orthogonal operators and hence spans \textcolor{blue}{the operator space} of the d- dimensional Hilbert space.\\

Within  the dual space, DAPG, the d(d+1) "points" are, naturally, (m,b) - b
associated with the (MUB) basis and m , vector within the basis. Note that the
lines, $L^{t},\;t=c,r,s$   may be parametrized for fixed c,r and s by
$\ddot{m}, m_0$, the vectors in the $b=\ddot{0}$ and $b=0$ that are on the
line. Thus DAPG provides a mapping scheme for (finite dimensional) Hilbert
space operators to (finite dimensional) phase space function. In particular it
allows a "natural" derivation of the (finite dimensional) Wigner function
\cite{wootters4,klimov1,khanna}. To this end we now show that the line
operators form a $d^2$ dimensional
orthogonal basis that may be used to express, e.g., the density operator.\\
Using Eq(\ref{lop}) we write, t=c,r,s ,\\
\begin{equation}\label{AB}
tr\hat{L}^{(t)}_{j=(\ddot{m},m_0)}\hat{L}^{(t)}_{j{'}=(\ddot{m}{'},m{'}_0)}=A+B+C+D+E.
\end{equation}
\begin{eqnarray}
A\;&=&\;\sum_{b}tr|m(b),b\rangle\langle b,m(b)|m{'}(b),b\rangle\langle b,m{'}(b)|=\begin{cases} 1\;\;\;
j\ne j{'}\\
d+1\;\;j=j{'}\end{cases} \nonumber\\
B\;&=&\;\sum_{b\ne b{'}}tr|m(b),b\rangle\langle b,m(b)|m{'}(b{'}),b{'}\rangle\langle b{'},m{'}(b{'})|
=d+1\nonumber \\
C\;&=&\;-\sum_{b}tr|m(b),b\rangle\langle b,m(b)|= -(d+1)\nonumber \\
D\;&=&\;-\sum_{b{'}}tr|m{'}(b{'}),b{'}\rangle\langle b{'},m{'}(b{'})|=-(d+1)\nonumber \\
E\;&=&\;\;tr\;\mathbb{I}=d
\end{eqnarray}
The value of A obtains, because for $j\ne j{'}$ within DAPG, the lines do have one common point: cf.
$\tilde{\lambda}_1,$ Section I. The value of B obtains, since we deal with MUB the scalar
product squared of any two states of different bases gives 1/d. There are d(d+1) terms in the sum thus
we have B=d+1 as stated.\\
We have thus,
\begin{equation}\label{orth}
tr\hat{L}^t_{j=(\ddot{m},m_0)}\hat{L}^t_{j{'}=(\ddot{m}{'},m{'}_0)}=d\delta_{\ddot{m},\ddot{m}{'}}
\delta_{m_0,m{'}_0}.
\end{equation}

However
\begin{equation}
\big(\hat{L}^t_{j=(\ddot{m},m_0)}\big)^2=\mathbb{I}\;\;only\;for\;r=1,\;s=0.
\end{equation}

\textcolor{blue}{It is only with the choice of "symmetric" line equation
family, r=1, s=0 in Eq.(\ref{lambda}), that  the (displaced) parity operator
is introduced in the mapping of Hilbert space formalism onto quantum mechanics
of phase space.}

\section{Finite Dimensional Phase space}

The line operators, $\hat{L}_{j=(\ddot{m},m_0)};\;\ddot{m},m_0=0,1,...d-1,$
was shown, within DAPG, Eq.(\ref{real}), to form an orthogonal $d^2$
dimensional set. We adopt an intuitively appealing view and consider
$\ddot{m}$, which is associated with the eigen value of the accounting
operator, Z, Eq.(\ref{z}), as designating position, while associating  $m_0$,
that relates  to its Fourier transform, with the momenta. Thus
$(\ddot{m},m_0)$ is viewed as a point in finite dimensional
phase space.\\

It was argued in the previous section that within APG, coordinated with $\ddot{m}$ along the positive x
direction and $m_0$
along the (positive) you axis, the \textcolor{blue}{DAPG {\it line}} operator,  $\hat{L}_{j=
(\ddot{m},m_0)}$,
is underpinned with \textcolor{blue}{APG {\it point}} $(\ddot{m},m_0)$.\\

a. Finite Dimensional Wigner Function.\\

Arbitrary Hilbert space operator, $\hat{A}$,  may be expanded in terms of the $d^2$ orthogonal
line operators,
\begin{equation}
\hat{A}=\frac{1}{d}\sum_j\big(tr \hat{A}\hat{L}_j\big)\hat{L}_j.
\end{equation}
$\big(tr \hat{A}\hat{L}_j\big)$ may be viewed within APG, wherein $j=(\ddot{m},m_0)$ is a point, as
a (finite dimensional) phase space representation of the (finite dimensional) Hilbert space operator,
$\hat{A}$:
$$\hat{A}\;\Rightarrow \;A(\ddot{m},m_0)=\big(tr \hat{A}\hat{L}_j\big).$$

Considering, in particular, the case $\hat{A}=\hat{\rho}$, the density
operator,
\begin{equation}
\hat{\rho}\Rightarrow
\rho(\ddot{m},m_0)=tr\hat{\rho}\hat{L}_{j=\ddot{m},m_0}\equiv
dW(\ddot{m},m_0).
\end{equation}
$W(\ddot{m},m_0)$ is the finite dimensional Wigner function, where the analogy is $\ddot{m}\sim q;\;m_0\sim p.$
This function is normalized
\begin{equation}
\sum_{\ddot{m},m_0}W(\ddot{m},m_0)=tr\hat{\rho}\big[\frac{1}{d}\sum_j\hat{L}_{j=(\ddot{m},m_0)}\big]=tr\rho=1.
\end{equation}
It is real,
\begin{equation}
W(\ddot{m},m_0)^{\ast}=\frac{1}{d}\big(\sum_{n,n'}
\langle n|\rho|n{'}\rangle\delta_{n{'}+n,2\ddot{m}}\omega^{-(n{'}-n)m_0}\big)^{\ast}
=\frac{1}{d}\sum_{n{'},n}\langle n|\rho|n{'}\rangle\delta_{n+n{'},2\ddot{m}}\omega^{-(n{'}-n)m_0}
=W(\ddot{m},m_0).
\end{equation}
It plays the role of a distribution,
\begin{equation}
tr\rho\hat{A}=\frac{1}{d^2}\sum_{j,j{'}}\big(tr\rho\hat{L}_j\big)\big( tr\hat{A}\hat{L}_{j{'}}\big)
\big(tr\hat{L}_j\hat{L}_{j{'}}\big)=
\sum_{\ddot{m},m_0}W(\ddot{m},m_0)A(\ddot{m},m_0).
\end{equation}
However the (finite dimensional) Wigner function is not positive definite. Thus consider two orthogonal
density matrices, $\rho_1, \rho_2$,
\begin{equation}
0=Tr\rho_1\rho_2=\frac{1}{d}\sum_{j=\ddot{m},m_0}\big(tr\rho_1\hat{L}_{j=\ddot{m},m_0}\big)
\big(tr\rho_1\hat{L}_{j=\ddot{m},m_0}\big)=d\sum_{\ddot{m},m_0}W_1(\ddot{m},m_0)W_2(\ddot{m},m_0)=0,
\end{equation}
implying that finite dimensional Wigner function is not positive definite and hence is "quasi distribution",
in close analogy with the Wigner function within the continuous phase space, \cite{ulf}.\\

b. Finite Dimensional Radon Transform.\\

We now review briefly some elements of phase space representation of Hilbert space operators in the
continuum,
$(d\Rightarrow \infty).$ This will guide us in our finite dimensional Radon transform formulation \cite{r1}.\\
Consider, within the continuum  the operator,
\begin{equation}
\hat{X}_{\theta}=\hat{x}C+\hat{p}S;\;\;C=cos\theta,\;S=sin\theta,\;\;\hat{x},\hat{p}
\;position\;and\; momentum\; operators\; resp.
\end{equation}
Denote its ($\delta$ function) orthonormalized eigenfunctions,
$|x;\theta\rangle$,
\begin{eqnarray}\label{cnmub}
\hat{X}_{\theta}|x{'};\theta\rangle&=&x{'}|x{'};\theta\rangle,\;\;\langle
\theta;x"|x{'};\theta\rangle=
\delta(x"-x{'}),\nonumber \\
\langle x|x{'};\theta\rangle=\frac{e^{\frac{i}{2S}\big[(x^2+x{'}^2)C-2xx{'}\big]}}{\sqrt{2\pi|S|}}.
\end{eqnarray}
The phase of the x representative wave function, $\langle x|x{'};\theta\rangle$, was chosen \cite{wootters4}
 to assure (for $-\infty \le x,x{'}\le \infty,\;0\le \theta\le \pi$)
\begin{equation}
lim_{\theta\rightarrow 0}\langle x|x{'};\theta\rangle=\delta(x-x{'});\;\;\;lim_{\theta\rightarrow \pi/2}
\langle x|x{'};\theta\rangle=\frac{e^{-ixx{'}}}{\sqrt{2\pi}}.
\end{equation}
The bases, $|x;\theta\rangle,\;|x{'};\theta{'}\rangle,\;for\;\theta\ne\theta{'}$ form an MUB sets
\cite{durt}:
\begin{equation}
|\langle \theta{'};x{'}|x{"};\theta{"}\rangle|=\frac{1}{2\pi|S(\theta{'}-\theta{"})|}\;\;independent
\;of\;x{'},x{"}.
\end{equation}

The Moyal \cite{moyal} mapping of the continuous Hilbert space operators,
$\hat{A},$ onto the continuous phase space, is given by
\begin{equation}
\hat{A}\Rightarrow A(q,p)=\int dq e^{-ipy}\langle q-y/2|\hat{A}|q+y/2\rangle.
\end{equation}
\textcolor{blue}{For $\hat{A}=\hat{\rho}$, the density operator, the RHS is
$2\pi W(q,p)$, where W(q,p) is the Wigner
function and the $2\pi$ is introduced to have it normalized to unity.}\\
Reverting to our notation, the Moyal transform of $\hat{A}$ reads,
\begin{equation}
\hat{A}\Rightarrow A(q,p)=tr \hat{A}\hat{L}_{q,p};\;\;\Rightarrow \langle x|\hat{L}_{q,p}|x'\rangle=
\delta\big(\frac{x+x'}{2}-q\big)e^{-ip(x-x')}.
\end{equation}
The Moyal transform of $\hat{A}=\hat{\mathbb{P}}(x;\theta)=|x;\theta \rangle \langle \theta;x|$ is,

\begin{eqnarray}\label{psmub}
\hat{\mathbb{P}}(x;\theta)\rightarrow\mathbb{P}(q,p)&=&\int e^{-ipy}\langle q-y/2|x;\theta\rangle
\langle \theta;x|q+y/2\rangle =\nonumber \\
tr\hat{\mathbb{P}}(x;\theta)\hat{L}_{(q,p)} &=&\delta(x-qC-pS),
\end{eqnarray}
where we used Eq.(\ref{cnmub}).\\
The Moyal transform abides by the {\it overlap formula}, \cite{ulf},
\begin{equation}
tr \hat{A}\hat{B}=\frac{1}{2\pi}\int dqdp\; tr\big(\hat{A}\hat{L}_{(q,p)}\big)tr\big(\hat{A}\hat{L}_
{(q,p)}\big)
=\int dqdp A(q,p)B(q,p),
\end{equation}
suggesting that the  Radon transform of Wigner function which is \cite{ulf,
wootters1, pier,rev2}
\begin{equation}\label{rd}
R[W](x;\theta)\equiv \tilde{\rho}(x,\theta)=\int dqdp W(q,p)\delta(x-qC-pS),
\end{equation}
allows  interpreting  Eq.(\ref{rd}) by: The Radon transform is the phase space
map of an MUB projector of the density operator. Carrying this to the finite
dimensional case, the projector of an MUB state mapped onto (finite
dimensional) phase space corresponding to the continuum expression,
Eq.(\ref{psmub}), is
\begin{equation}
\hat{S}_{\alpha=(m;b)}\equiv\hat{\mathbb{P}}(m;b)\;\rightarrow\;\mathbb{P}(\ddot{m},m_0)
= tr \hat{S}_{\alpha=(m;b)}\hat{L}_{j=(\ddot{m},m_0)}\equiv
\Lambda_{\alpha,j}=\begin{cases} 1,\;\alpha\in j\\0,\;\alpha \notin
j.\end{cases}
\end{equation}

Thence  the finite dimensional Radon transform (of Wigner function) is (cf.
Eq.(\ref{rd},\ref{psmub}))
\begin{eqnarray}\label{Lambda}
R[W](m,b)&=&tr\hat{\rho}\hat{S}_{\alpha};\;\;\alpha=(m,b);\;j=(\ddot{m},m_0);
\nonumber \\
&=&\frac{1}{d}\sum_{j=\ddot{m},m_0} tr \hat{\rho}\hat{L}_j tr
\hat{\mathbb{P}}(m,b)\hat{L}_j \nonumber \\
&=&\sum_{\ddot{m},m_0}W(\ddot{m},m_0)\Lambda_{\alpha,j}.
\end{eqnarray}
where $\Lambda$ plays the role of the $\delta$ function in the continuum.\\
Generalizing the Radon transform of an arbitrary finite dimensional phase
space function, $Q(\ddot{m},m_0)$,
\begin{eqnarray}
\hat{Q}\Rightarrow
Q(\ddot{m},m_0)&=&\frac{1}{d}tr \hat{Q}\hat{L}_{\ddot{m},m_0}. \nonumber \\
 R[W](m,b)&=&tr\hat{Q}\hat{S}_{\alpha=(m,b)}=\sum_{\ddot{m},m_0}Q(\ddot{m},m_0)\Lambda_{(\ddot{m},m_0),(m,b)}.
\end{eqnarray}

c. Phase Space Mappings of Finite Dimensional Wave Functions.\\

We now consider DAPG underpinning of a two d-dimensional particles {\it wave
function}, $|m;b\rangle_1|\tilde{m};\tilde{b}\rangle_2$ (the notation are
defined in section II)\cite{r4}: The (m,b) coordinate scheme for the DAPG now
underpins the two particles wave function rather than
the operator $\mathbb{P}(m,b)$ considered above.\\

Underpinning two d-dimensional particles product MUB states with geometrical
points, we get, as the definition of the corresponding DAPG "line state",
\begin{eqnarray}
S_{\alpha=(m,b)}&\Rightarrow& |S_{\alpha=(m,b)}\rangle
=|m;b\rangle_1|\tilde{m};\tilde{b}\rangle_2 \nonumber \\
&\Rightarrow&|m;b\rangle_1|\tilde{m};\tilde{b}\rangle_2=\frac{1}{d}\sum_{j\in
\alpha=(m,b)}|L_{j}\rangle.
\end{eqnarray}

Inverting, i.e. expressing the "line states" in terms of the "point states",
cf Eq.(\ref{real}), gives
\begin{equation}
|L_j\rangle=\sum_{\alpha\in j}|S_{\alpha}\rangle -|{\cal{R}}\rangle;\;\;|{\cal{R}}\rangle=
\frac{1}{d+1}\sum_1^{d(d+1)}|S_{\alpha}\rangle=\frac{1}{d}\sum_1^{d^2}|L_j\rangle.
\end{equation}
The universal (wave) function $|{\cal{R}}\rangle$ is
\begin{equation}\label{unfn}
|{\cal{R}}\rangle=\frac{1}{d+1}\sum_{(m,b)}^{d(d+1)} |m;b\rangle_1|\tilde{m};\tilde{b}\rangle_2=
\sum_{m}|m;b\rangle_1|\tilde{m};\tilde{b}\rangle_2.
\end{equation}
The RHS is gotten by noting that
\begin{equation}
\sum_{m}|m;b\rangle_1|\tilde{m};\tilde{b}\rangle_2=\sum_{m{'}}|m{'};b{'}\rangle_1|\tilde{m{'}};\tilde{b{'}
}\rangle_2.
\end{equation}
i.e. the summation over pencil states is independent of the pencil proving Eq.
(\ref{unfn}).\\

Thus the geometrical considerations gives for the "line" state,
($\alpha=(m,b);\;j=(\ddot{m},m_0)$), the expression
\begin{eqnarray}
|L_{j=(\ddot{m},m_0)}\rangle&=&\sum_{b}|m(b);b\rangle_1|\tilde{m}(b);\tilde{b}\rangle_2-
\sum_{m}|m;b\rangle_1|\tilde{m};\tilde{b}\rangle_2 \nonumber \\
&=&\sum_{n,n{'}}|n\rangle_1|n{'}\rangle_2\big[\sum_b\langle n|m(b);b\rangle\langle n{'}
|\tilde{m}(b);\tilde{b}\rangle - \sum_{m}\langle n|m;b\rangle \langle n{'}
|\tilde{m};\tilde{b}\rangle\big]\nonumber \\
&=&\sum_{n,n{'}}|n\rangle_1|n{'}\rangle_2\big[\langle n|\big(\sum_b
|m(b);b\rangle\langle b;m(b)|
-\mathbb{I}\big)|n{'}\rangle\big] \nonumber \\
&=&\sum_{n,n{'}}|n\rangle_1|n{'}\rangle_2\delta_{n+n{'}-2\ddot{m}}\omega^{-(n-n{'})m_0} \nonumber \\
&=&\omega^{2\ddot{m}m_0}\sum_{n}|n\rangle_1|2\ddot{m}-n\rangle_2\omega^{-2m_0n}.
\end{eqnarray}
Where we used the results of Section III. The last expression for the line state is evidently a
(not normalized) MES. Written in terms of the collective coordinates , Eq.(\ref{relcom}), and normalizing
we write
\begin{eqnarray}\label{ent1}
\frac{1}{\sqrt {d}}|L_{j=(\ddot{m},m_0)}\rangle&=&\frac{\omega^{2\ddot{m}m_0}
|\ddot{m};\ddot{0}_c\rangle_c}{\sqrt{d}}\sum_{n}
|n-\ddot{m};\ddot{0}_r\rangle_r\omega^{-2m_0n} =
|\ddot{m};\ddot{0}_c\rangle_c|2m_0;0_r\rangle_r.\;\;\Rightarrow \nonumber \\
\frac{1}{\sqrt
{d}}|L_{0;0}\rangle&=&|0;\ddot{0}\rangle_c|0;0\rangle_r\;\;\;origin\;of\;"phase\;space".
\Rightarrow \nonumber \\
\hat{X}^{\ddot{m}}_c\hat{Z}^{-2m_0}_r\frac{1}{\sqrt
{d}}|L_{0;0}\rangle&=&\frac{1}{\sqrt {d}}|L_{j=(\ddot{m},m_0)}\rangle.
\end{eqnarray}
The MUB sets $b=\ddot{0}$  (the computational basis (CB)) and $b=0$, its
Fourier transform basis are, of course, complete orthonormal bases. Thence the
"line" states $|L_{j=(\ddot{m},m_0)}\rangle$ form a $d^2$ orthogonal MES basis
that spans the $d^2$ dimensional two particle states:
\begin{equation}
\langle L_{\ddot{m}{'},m_0{'}}|L_{\ddot{m},m_0}\rangle=\delta_{\ddot{m}{'},\ddot{m}}\delta_{m_0{'},m_0},\;\;
\ddot{m}{'},\ddot{m},m_0{'},m_0=0,1,2...d-1.
\end{equation}
This orthogonality may be proved  via the approach of Eq.(\ref{AB}).\\

A conjugate MES basis is,
\begin{eqnarray}\label{conj}
|\tilde{L}_{m_0;\ddot{m}}\rangle&=&|2m_0;0\rangle_c|\ddot{m};\ddot{0}\rangle_r. \Rightarrow\nonumber \\
\langle\tilde{L}_{m_0;\ddot{m}}|L_{\ddot{m}{'},m_0{'}}\rangle&=&\langle
0;2m{'}_0|_c\langle
\ddot{0};\ddot{m}{'}|_r|\ddot{m};\ddot{0}_c\rangle_c|2m_0;0_r\rangle_r=
\frac{1}{d}\omega^{-2m_0\ddot{m}{'}}\omega^{2m{'}_0\ddot{m}}.
\end{eqnarray}
Note that the $d^2$ dimensional orthonormal MES basis,
$|2m_0;0\rangle_c|\ddot{m};\ddot{0}\rangle_r;\;\;m_0,\ddot{m}=0,1,2,...d-1$
forms a MUB to the $d^2$ dimensional MES basis,
$|2m_0;0\rangle_r|\ddot{m};\ddot{0}\rangle_c$:
\begin{equation}
|\langle \tilde{L}_{m_0;\ddot{m}}|L_{\ddot{m}{'},m_0{'}}\rangle|=\frac{1}{d}.
\end{equation}

d. Mean King Problem (MKP) and Tracking the Mean King (TMK).\\

\noindent A DAPG underpinning of an Hilbert space allows a direct solution to the MKP as well as its variant,
TMK. \\
The MKP is a quantum mechanical retrodiction problem \cite{mermin,werner,a2}.
It was posed originally in \cite{lev} for spin 1/2 particles, extended in
\cite{berge} to prime dimensionality, d=p, and to powers of primes in
\cite{aravind,durt1,durt}. Further generalization are discussed in
\cite{werner}. We consider in the following the d=p cases.\\
The MKP (i) and TMK (ii) involves a (two particles) state prepared by Alice.
One of the particles is availed to the King who measures its state in an MUB,
b, of his choice. Subsequent to his measurement, Alice performs a control
measurement of the two particle state. Now  i. Within the MKP, Alice is
challenged to infer the outcome of the King's measurement, say, m when,
\textcolor{blue}{after} she completes her control measurement, she is told the
basis, b,  used by the King in his measurement. ii. Within the TMK she is
challenged to deduce , via her control measurement,
 the basis used by the King. \\
We consider sequentially the MKP and TMK.  \\
Let Alice prepare the universal "line state",Eq.(\ref{unfn}),
\begin{equation}\label{is}
|{\cal{R}}\rangle=\sum_{m}|m;b\rangle_1|\tilde{m};\tilde{b}\rangle_2 \nonumber
\end{equation}
The King measures, $\hat{K}$,
\begin{equation}
\hat{K}=\sum_{m'}|m{'}:b\rangle K_{m'}\langle b;m{'}|_1,
\end{equation}
for some b of his choice with an outcome, say, m.\\
i. To handle the MKP, Alice measures $\hat{\Gamma}$ for her control measurement.\\
\begin{equation}\label{gamma1}
\hat{\Gamma}=\sum_{\ddot{m},m_0}|\ddot{m}\rangle_c|2m_0;0\rangle_r\Gamma_{\ddot{m},m_0}
\langle \ddot{m}|_c\langle 0;2m_0|_r,
\end{equation}
with outcome, say $m{'}_0,\ddot{m}{'}$.\\
Thence,

\begin{equation}
 \langle \ddot{m}{'}|_c\langle 2m{'}_0;0|_r |m;b\rangle_1|\tilde{m};\tilde{b}\rangle_2\ne 0.
\end{equation}
The matrix element gives the probability amplitude for the "point state"
$|m;b\rangle_1|\tilde{m};\tilde{b}\rangle_2$ to be on the "line state"
$\langle \ddot{m}{'}|_c\langle 2m{'}_0;0|_r$. This is non nil only if the
geometrical point (m,b) is on the geometrical line viz
\begin{equation}
m=m{'}_0+b\ddot{m}{'}-b/2 ;\;\;b\ne \ddot{0}.
\end{equation}
For $b=\ddot{0},$ evaluation of the matrix element gives, $m=\ddot{m}{'}$.
Since Alice knows $m{'}_0,\ddot{m}{'},$ she may infer the value of m upon
being \textcolor{blue}{informed} of the value of b. \\

ii. To deal with TKM, Alice measures $\hat{\tilde{\Gamma}}$ for her control
measurement. With
$\hat{\tilde{\Gamma}}$   conjugate to $\hat{\Gamma}$, thus,\\
\begin{equation}\label{gamma1}
\hat{\tilde{\Gamma}}=\sum_{\ddot{m},m_0}|\ddot{m}\rangle_r|2m_0;0\rangle_c
\tilde{\Gamma}_{\ddot{m},m_0} \langle \ddot{m}|_r\langle 0;2m_0|_c,
\end{equation}
with outcome, say, $m{'}_0,\ddot{m}{'}$.\\
We have then,
\begin{equation}
\langle\ddot{m}{'}|_r\langle 2m{'}_0;0|_c
|m;b\rangle_1|\tilde{m};\tilde{b}\rangle_2 \ne 0.
\end{equation}
The non vanishing of this matrix element implies,
\begin{equation}
b=\begin{cases} -\frac{m{'}_0}{\ddot{m{'}}};\;\ddot{m}{'}\ne 0.\\
\ddot{0};\;\;m{'}_0\ne 0,\;\ddot{m}{'}=0.\\undetermined\;\;\ddot{m}{'}=m{'}_0=0. \end{cases}
\end{equation}
Since Alice knows  $m{'}_0,\ddot{m}{'},$  she can deduce b, the basis
used by the King, except for the case, whose probability is $1/{d^2}$, when $m{'}_0=\ddot{m}{'}=0$,
that leaves b undetermined.\\

Note:\\

\noindent 1. In the special case, whose probability is $1/{d^2}$, that her control
measurement yields back the initially prepared line:
i.e. $m{'}_0=\ddot{m}{'}=0$, \textcolor{blue}{Alice does not gain any information.} \\

\noindent 2. Alice's deduction is independent of the out come of the king's
measurement. His measurement, in effect,  non selective.\\

\section{Summary and Concluding Remarks}

Finite affine plane geometry (APG) and its dual, DAPG, were studied within the
general theory of linear spaces. DAPG was found most convenient for our study
and we consider lines and points within the theory. (The basic notions of
linear spaces, APG and DAPG are given in Section II. We refer to the order of
a geometry as its dimension to simplify the notation.) It was shown that for a
d-dimensional Hilbert space the existence of d+1 mutually unbiased bases (MUB)
(MUB are introduced in Section III) implies the existence of d-dimensional
DAPG. (Which, in turn, implies the existence of APG as well as projective
geometry.) DAPG is used to define lines and points {\it operators} that act in
the overriding Hilbert space. The interrelation among these operators
is determined by the underpinned DAPG.\\

 Inversion of the expression that gives  the "point operator" in terms of  "line operators", requires
 the existence of a geometrically based universal operator, i.e. one that is independent of
of either points or lines. Expressing operators (and states) via mutually
unbiased bases (MUB) lead to simple universal operators. Thus having the
geometrical points underpin MUB projectors provides for unity as the universal
operator.\\

The "line operators" are shown to form a $d^2$  dimensional orthogonal basis
for a d dimensional Hilbert space \textcolor{blue}{operators}. These
 line operators provides a convenient means for the introduction of
finite dimensional phase space. Thus representations of Hilbert space
operators in terms of these line operators may thus be interpreted as mappings
of q-number Hilbert space entities onto c-number phase space functions. An
example of such function is, as is well known, the Wigner function viewed here
as a mapping of $\rho$, the density matrix, onto (finite dimensional) phase
space wherein
 $\rho$ is expressed in terms of the "line operators".\\

The definition of the geometries as constrained linear space underscores their
non unique lines constituency. (These notions are introduced in Section II.)
The existence of several distinct line formation for a d dimensional DAPG
implies that there exist several distinct sets of Hilbert space "line
operators" for a d dimensional Hilbert space. This, in turn, leads to distinct
mappings of the Hilbert space onto (finite dimensional) phase space. For
example, it is shown that only one such
possible map relates to the celebrated Wigner function connection with the parity operator.\\

Underpinning two particle MUB \textcolor{blue}{product} state with DAPG points gives as "line states"
maximally entangled states (MES). The $d^2$ orthogonal "line states" of this case form a complete
MES orthonormal basis for the two particle Hilbert space. The universal state implied here allows
a concise
solution to the so called Mean King Problem.\\

The correspondence of finite dimensional line operators within dual affine plane geometry, DAPG,
with finite
dimensional point operators affine plane geometry, APG,  (and vice versa) is established. Thus confirming
the basic equivalence of using either for mapping of Hilbert space q functions to phase space c functions.\\

The d(d+1) points of a d-dimensional DAPG were used also to underpin d(d+1)  two particles MUB {\it product
states}. In this case the so called "line states" $|L_j\rangle$ underpinned with the corresponding
DAPG lines are
maximally entangled states (MES) of the two particles. These in turn were shown to be product states when
accounted for in terms of collective coordinates. The universal function implied in this case allowed a
concise solution to the so called Mean King problem and its extension.\\

The present work underscores the fundamental role played by Mutually Unbiased Bases (MUB) in relating
physical and
geometrical entities and provides a geometrical approach for a phase space formulation of quantum mechanics.\\

\section*{Appendices }

The following appendices contains sample proofs aimed at  illustrating  some of the mathematics
as well as provide examples that hopefully clarifies some perhaps involved mathematical ideas.\\

A. Extension of GF(3) to $GF(3^2).$\\

 We now  illustrate how the extended field $GF(p)\;\Rightarrow\;GF(p^m),$ m
a
positive integer allows the construction of APG. Our example is $d=3^2$.\\
Consider a polynomial of degree 2 (m=2) with coefficients in GF(3) (p=3)
\textcolor{blue}{that has no roots in GF(3)}. e.g. (-1=2 Mod[3])
\begin{equation}
q(u)=u^2-u-1.
\end{equation}
(One can readily check that no $u\in GF(3),$ (viz. u=0,1,2) is a root.) The
 field is now extended to include the root of q(u), viz $u\in q(u)=0$. Thus
 the elements of the field are the elements of GF(3) and their sum and product
 with u - the root of the irreducible polynomial, q(u):
\begin{equation}
GF(3^2)=\{0,1,2,u,u+1,u+2,2u,2u+1,2u+2\}.
\end{equation}
For example:
$$(2u+1)+(2u+2)=u;\;\; (2u+1)(2u+2)=u;\;\;\frac{1}{2u+1}=2u.$$
We now consider a square, 9x9, array whose horizontal direction (x axis) is
labelled with
$$x\;=\;0,1,2,u,u+1,u+2,2u,2u+1,2u+2,$$
with similar labelling for the vertical direction (y axis).  A "point" is the
"coordinate" (x,y) and the lines $L_{(\alpha,\alpha')},$ $((\alpha, \alpha')$
two distinct points) are given, for 81 $(3^2\cdot3^2)$ of them by a linear
relation,
\begin{equation}
y\;=\;ax\;+\;b;\;\;\;x,y,a,b\;\in\;GF(3^2)
\end{equation}
and the other 9 $(3^2)$ by
\begin{equation}
x=b;\;\;x,b\;\in\;GF(3^2).
\end{equation}
Thus we have $d^2=81$ points and $d(d+1)=90$ lines with $d+1=10$ pencils (i.e.
sets of parallel lines) thence we have constructed a FAPG. We are thus assured
that for this dimensionality both DAPG and PPG exist.\\

B. Proof that in a linear space knowing the number of points is $d^2$ and that the number of points per line
is d, the same for all lines suffice to show that the linear space is dual affine plane geometry, DAPG.\\

Given  a linear space \cal{S} wherein the following items holds i. $
\nu=d^2$, and ii. $k_{L_j}=k_L=d$. We then prove that
\cal{S}=\cal{A} i.e.  DAPG.\\
We first show i. and ii. implies b=d(d+1) and $r_{S_{\alpha}}=r_S=d+1$:\\
Consider an arbitrary point, $S_{\beta}$. We now count the points residing on the $r_{S_{\alpha}}$
lines that share the point $S_{\beta}$ excluding the point itself. This number is, given ii above, is
$(d-1)r_{S_{\beta}}$. Since  $S_{\beta}$ is connected uniquely, given $\lambda 1$, to every other point,
all points are counted and each point is counted once. Thus we have using i.,
\begin{equation}
(d-1)r_{S_{\beta}}=\nu-1=d^2-1 \Rightarrow \;r_{S_{\beta}}=r_S=d+1.
\end{equation}
Now counting incidences in two ways \cite{batten} (p. 14), we have quite generally, Eq.(\ref{univ}),
\begin{equation}
\sum_{\alpha}r_{S_{\alpha}}=\sum_{j}k_{L_j}\;\Rightarrow \;r_S \nu =k_Lb\;\Rightarrow \; (d+1)d^2=db.
\end{equation}
i.e. b=(d+1)d. QED
Now $r_S=d+1$ and $k_L=d$ implies A: $\lambda 1$ implies that d lines on $S_{\alpha}$ not on $L_j$
connects it with $L_j$ and exactly one that does not.\\

C. Properties of Pencils within DAPG. \\

Generalities: $r_s$=number of pencils, p. $r_{cs}$=number of points in DAPG pencil.\\
i. Pencils are mutually exclusive.\\
Let $\Pi$ define pencil relation: $\alpha\Pi \alpha{'}\Leftrightarrow \alpha\; and\; \alpha{'}$ are not connected by a line. \\
We shall show that $\alpha\Pi \alpha{'},\;\alpha\Pi \alpha{"}\Rightarrow\alpha{'}\Pi \alpha{"}$.\\
Proof.\\
Suppose $\alpha{'}$ and $\alpha"$ are joined by a line, $L(\alpha{'},\alpha")$. This line has \textcolor{blue}{two}
points not connected to $\alpha$. Contradiction (note axiom $\tilde{\cal{A}}$.\\
Thus a pencil is an equivalence class \cite{bennett}.\\

ii. Pencils are equally populated ($r_{cs}$).\\
Consider a point $S_{\alpha}$. A set of d lines share it. Consider a line $L_j{'}$  not in this set. It has
one point, $S_{\alpha{'}}$, not connected to $S_{\alpha}$ (cf. Axiom $\tilde{\cal{A}}$). There are d lines sharing through this point. Since there are all together $d^2$ lines there are d-1 such sets each with d lines and each with a point $S_{\alpha"}$ not connected to $S_{\alpha}$. With $S_{\alpha}$ there are thus d point in a pencil, $r_{cs}=d$.\\
iii. There are $r_s=k_L$ pencils.\\
There are d+1 points on a line, each belonging to a
distinct pencil. Each pencil is an exclusive set containing d points. There
are all together d(d+1) points. Thus d(d+1)=d($r_s$) $\Rightarrow \;r_s=d+1=k_L.$\\

iv. The universal function and pencil's constituents.\\
We now prove that The definition, Eq.(\ref{real}):
$$\hat{S}_{\alpha}=\frac{1}{d}\sum_{j\in \alpha}\hat{L}_j\Rightarrow \sum_{\alpha\in p}\hat{S}_{\alpha}=
 \sum_{\alpha{'}\in p{'}}\hat{S}_{\alpha{'}}.$$ i.e. the RHS sum is independent of the pencil, p.\\
Proof:\\
\begin{eqnarray}
\hat{S}_{\alpha}&=&\frac{1}{d}\sum_{j\in \alpha}\hat{L}_j\Rightarrow \nonumber \\
\sum_{\alpha\in p}\hat{S}_{\alpha}&=&\frac{1}{d}\sum_j^{d^2}\hat{L}_j,\Rightarrow\nonumber \\
\sum_{\alpha\in p}\hat{S}_{\alpha}&=&\sum_{\alpha{'}\in p{'}}\hat{S}_{\alpha{'}}. \;\;QED.
\end{eqnarray}

D. Proof of Eq.(\ref{shiftinversion}).\\

Consider $|m(b);b\rangle$ for $m(b)=m_0+b\ddot{m}-b/1,\;\;for b\ne \ddot{0};$
 $m(\ddot{0})=\ddot{m}.$\\
For $\ddot{m},m_0=0$ the line made of the projectors $|-b/2;b\rangle\langle b;-b/2|$\\
We have,
$$|-b/2;b\rangle=\frac{1}{\sqrt d}\sum_{n=0}^{d-1}|n\rangle\omega^{\frac{b}{2}n^2}.$$
Thus,
\begin{eqnarray}
\hat{X}^{\ddot{m}}\hat{Z}^{-m_0}|-b/2;b\rangle&=&\frac{1}{\sqrt d}\sum
|n+\ddot{m}\rangle\omega^{\frac{b}{2}n^2-m_0 n}=\nonumber \\
=\frac{\omega^{\ddot{m}(\frac{b}{2}\ddot{m}+m_0)}}{\sqrt d}
\sum|n\rangle\omega^{\frac{b}{2}n(n-1)-(m_0+b\ddot{m}-b/2)n}
&=&\omega^{\ddot{m}(\frac{b}{2}\ddot{m}+m_0)}|m_0+b\ddot{m}-b/2;b\rangle.
\end{eqnarray}

Hence
\begin{equation}
\hat{X}^{\ddot{m}}\hat{Z}^{-m_0}\hat{L}_{0,0}\hat{Z}^{m_0}\hat{X}^{-\dot{m}}=\hat{L}_{\ddot{m},m_0}.
\end{equation}
QED.\\

\end{document}